# Explicit gain equations for hybrid graphene-quantum-dot photodetectors

Kaixiang Chen[1,#], Chufan Zhang[1,#], Xiaoxian Zang[2], Fuyuan Ma[2], Yuanzhen Chen[3]* and Yaping Dan[1]*

[1]University of Michigan – Shanghai Jiao Tong University Joint Institute, Shanghai Jiao Tong University, Shanghai 200240, China

[2]Key Laboratory of Solar Energy Utilization & Energy Saving Technology of Zhejiang Province, Zhejiang Energy R&D Institute Co., Ltd., Hangzhou 311121, China

[3]Institute for Quantum Science and Engineering and Department of Physics, Southern University of Science and Technology, Shenzhen 518055, China

# These authors contribute equally.

*Email: yaping.dan@sjtu.edu.cn, chenyz@sustech.edu.cn

**Abstract**

Graphene is an attractive material for broadband photodetection but suffers from weak light absorption. Coating graphene with quantum dots can significantly enhance light absorption and create extraordinarily high photo gain. This high gain is often explained by the classical gain theory which is unfortunately an implicit function and may even be questionable. In this work, we managed to derive explicit gain equations for hybrid graphene-quantum-dot photodetectors. Due to the work function mismatch, lead sulfide (PbS) quantum dots coated on graphene will form a surface depletion region near the interface of quantum dots and graphene. Light illumination narrows down the surface depletion region, creating a photovoltage that gates the graphene. As a result, high photo gain in graphene is observed. The explicit gain equations are derived from the theoretical gate transfer characteristics of graphene and the correlation of the photovoltage with the light illumination intensity. The derived explicit gain equations fit well with the experimental data, from which physical parameters are extracted.

Graphene is a zero-bandgap semimetal with extraordinarily high carrier mobility,[1,2,3,4,5] as a result of which graphene is an attractive material for broadband photodetection. Photodetectors based on graphene operating in the mid-infrared spectrum have been demonstrated in recent years.[6,7,8,9] However, due to its nature of being atomically thin, graphene suffers from weak light absorption, resulting in poor photoresponsivity.[10,11,12,13] Coating graphene with semiconducting quantum dots (QDs) can strongly enhance the light absorption and introduce an interesting high photo gain at an order of $10^8$,[14,15,16] several orders of magnitude larger than photodetectors based on pure semiconducting QDs (often have a photo gain of $10^2$-$10^3$).[17,18,19] The classical carrier-recycling gain mechanism is often used to explain the origin of high gain,[14,15,16] that is, the high gain originates from the photoexcited carriers circulating the circuits many times before recombination due to the long response time and short transit time.[20]

However, this classical gain theory is an implicit function and may even be questionable.[21] It is implicit in that it is a function of carrier lifetime and transit time and cannot quantitively fit the light-intensity-dependent photo gains. More importantly, the classical gain theory was derived on two questionable assumptions.[21,22] Firstly, the classical theory assumes no metal-semiconductor boundary confinement, which leads to the questionable conclusion that high gain can be obtained as long as the minority recombination lifetime is much longer than the transit time. After the metal-semiconductor boundary confinement is considered, it turns out that a photoconductor intrinsically has no gain or at least no high gain no matter how long the minority recombination time and how short the transit time is.[21] However, high gains in photoconductors are often observed in experiments. This is because the classical theory makes a second questionable assumption that the number of excess electrons and holes contributing to photoconductivity are equal.[21] Although excess electrons and holes are generated in pairs, excess minority carriers are often trapped by defects or potential wells in semiconductors. The same number of excess majority counterparts is accumulated in the conduction channel, leading to the experimentally observed high photogain.[21, 23]

After correcting these two assumptions, we further derived the explicit gain equations for single crystalline nanowires based on photo Hall measurements.[24] The derived gain equations are a function of light intensity and device physical parameters such as doping concentration, nanowire diameter and surface depletion width. The gain equations fit well with the experimental data from which we extracted parameters including minority carrier recombination lifetimes that are consistent with experimental results in literature.[24, 25] Although the photo gain is still proportional to the ratio of minority recombination lifetime to transit time, the explicit gain equations show that the high photo gain ($10^6$ - $10^8$) does not originate from this ratio since the ratio is not more than 10. Instead, the high photo gain comes from the light-illumination-induced photovoltage across the surface depletion region that modulates the conduction channel width.

Inspired by our previous work,[24] here we managed to derive explicit gain equations for hybrid graphene-quantum-dot photodetectors, which fit well with the experimental photoresponses. Due to the work function mismatch, the coating of QDs on graphene induces electron transfer from the QDs into graphene, which shifts the gate transfer characteristics of graphene and creates a depletion region in the QDs. The light illumination narrows down the depletion region width, creating a photovoltage across the graphene and QDs. The photovoltage gates graphene and thus induces a high photo gain in graphene observed in experiments. The explicit gain equations are derived from the theoretical gate transfer characteristics of graphene and the correlation of the photovoltage with light illumination intensity. The equations fit well with the experimental data, from which physical parameters are extracted.

**Results and Discussion**

Fig.1a shows the optical microscopic image of a graphene field effect transistor (GFET) fabricated on highly doped $SiO_2$/Si wafers with 1 μm thick $SiO_2$ following the procedure described below. A 150 nm thick Al gate electrode was first deposited on the $SiO_2$/Si wafer by photolithography and thermal evaporation. $HfO_2$ was then deposited on top of the Al gate by Plasma-Enhanced Atomic Layer Deposition (PEALD) at 250 °C. Five Au/Cr electrodes were formed near the gate electrode by a second time photolithography and thermal evaporation. As the next step, a single layer graphene sheet was transferred from a Cu foil to the sample surface in contact with the Au electrodes. Photolithography and oxidation plasma were applied in order to pattern the graphene into a Hall bar geometry. The schematic of the as-fabricated graphene device is shown in Fig.1b. As the last step, PbS quantum dots (QDs) with organic ligands were spin-coated on the GFET structure. See Experimental section for fabrication details. The measurements were undertaken in a vacuum chamber at a controlled temperature of 300K by placing the devices in a physical parameter measurement system (PPMS Evercool-II).

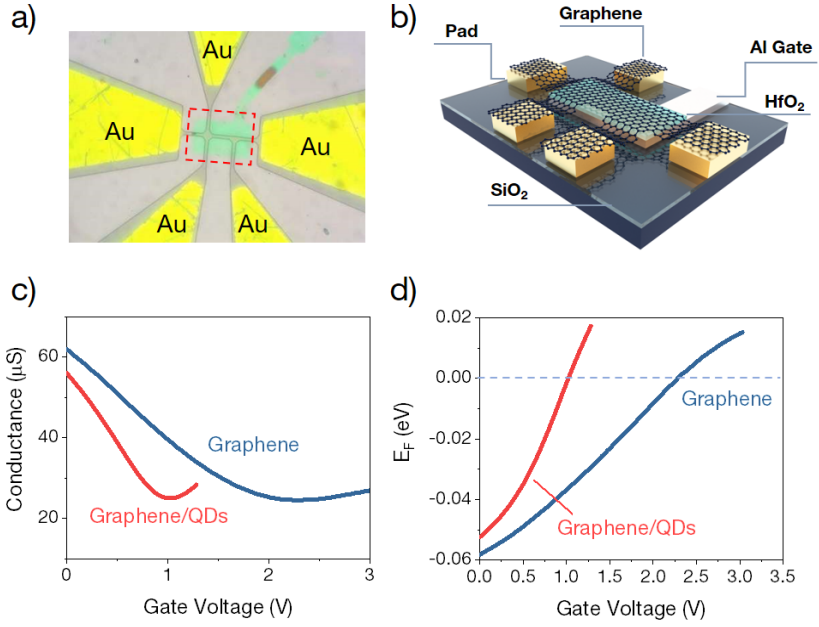

Fig. 1 a) Optical image of a graphene field effect transistor. b) Schematic of the as-fabricated graphene device. c) Gate transfer characteristics for graphene and graphene/QDs device in darkness at a bias of 0.5V between source and drain. d) Fermi energy level of graphene derived from the data in c).

Gate transfer characteristics of the graphene devices before and after coating with quantum dots are shown in Fig.1c. Before coating, the Dirac point is located at ~ 2.2V because of localized negative charges in surface states at graphene-$SiO_2$ interface that induce holes in graphene. At zero gate voltage, the graphene is p-type. Applying positive voltage on gate will push away holes in graphene and reduce the channel conductance until it reaches the minimal value at the Dirac point. After coating with quantum dots, the Dirac point is left-shifted to ~ 1V, similar to previous observations in literature.[26] The left-shift of the Dirac point is probably caused by the fact that these QDs have a smaller work function than graphene before QDs coating. Upon coating of QDs, electrons will transfer from QDs into graphene, neutralizing holes in graphene and making the graphene less p-type at $V_g$ = 0V (Fermi level moves closer to Dirac point). The gate voltage modulates the concentration of charge carriers and hence the conductance $\sigma$ of the graphene devices that is given by eq.(1).

$$\sigma = q \cdot \mu \cdot \frac{W}{L}(n + p) \quad (1)$$

where q is the unit charge, μ is the charge carrier mobility in graphene, W is the device width, L is the device length, n and p are the electron and hole concentration (per unit area) in graphene which are expressed as in eq.(2), respectively.[27]

$$n = \frac{n_i J_1(\frac{E_F}{kT})}{J_1(0)} \; ; \; p = \frac{n_i J_1(-\frac{E_F}{kT})}{J_1(0)} \quad (2)$$

in which $n_i$ is the intrinsic electron concentration (per unit area) of graphene, $E_F$ is the Fermi energy level respective to the Dirac point, k is the Boltzmann constant and T is the absolute temperature. $J_1(x) = \int_0^\infty \frac{u}{(1+e^{u-x})} du$ is the Fermi-Dirac integral.

Hall effect measurements were conducted in a dark vacuum chamber from which we found the mobility of charge carriers in graphene as ~ 2700 cm$^2$V$^{-1}$s$^{-1}$. The mobility remains nearly independent of the light illumination and gate voltage (less than 3% variation from 0V to 1.5V). At a given gate voltage we calculated the Fermi energy level from the corresponding conductance based on eqs.(1-2). The results are shown in Fig.1d. As the gate voltage moves positive, electrons flow into the graphene devices. Part of the electrons goes into the graphene conduction channel, shifting up the Fermi energy level from valence band and eventually into conduction band. As a result, the conductivity of the p-type graphene first reduces to a minimum value and then increases as the Fermi level moves across the Dirac point to conduction band. The remaining part of electrons fills the surface states below the Fermi level. These electrons do not contribute to the conductivity of graphene but gate the graphene device (right-shifting the Dirac point). If there are no surface states near graphene (but with fixed charges), the gate-induced electrons can only go into graphene conduction channel. In this case, the conductance of graphene at point "1" in Fig.2a will move to point "2" following the intrinsic gate transfer characteristics (dashed curve). However, because part of the electrons will actually fill the surface states and gate the graphene device, the intrinsic gate transfer characteristics will right shift (from "2" to "3") at the same time. The combined result of these two processes is that the appeared gate transfer characteristics moves from "1" to "3" following the solid black line. As the gate voltage increases, the intrinsic Dirac point V$_{dirac}$ also right shifts and eventually crosses the appeared Dirac point (red dot meets green dot in Fig.2a) where the Fermi energy level E$_F$, the appeared and intrinsic Dirac point are all aligned. Note that the movement of mobile charges is equivalent to filling surface states with charges.

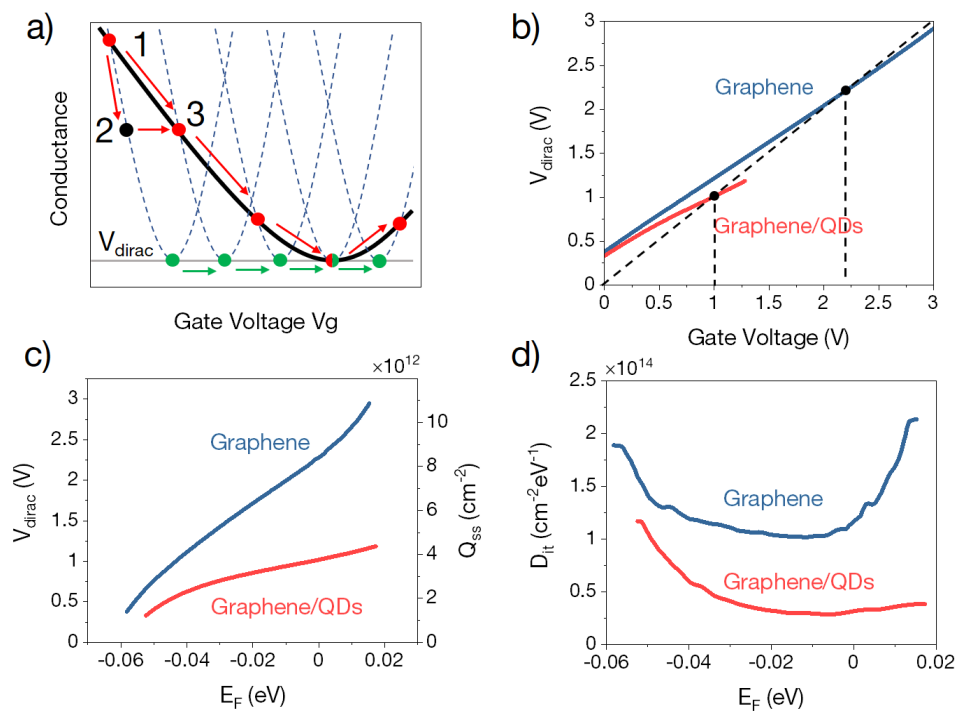

Figure 2. a) Intrinsic (dashed curves) and appeared gate transfer characteristics (solid curve). The appeared gate transfer characteristics is a combinational effect of intrinsic gate transfer characteristics and shifting of intrinsic Dirac point (green dots). b) Intrinsic Dirac point dependent on gate voltage. The black dots represent where the red and green dot in panel a meet. c) Intrinsic Dirac point dependent on Fermi energy level. d) Effective density of trap states derived from panel c.

Analytically, the appeared gate transfer characteristics is governed by eq.(3).[28, 29] Note that the intrinsic Dirac point $V_{dirac}$ is a function of $V_g$, representing the shift of the intrinsic gate transfer characteristics due to the filling of charges into localized states near graphene. $E_F$ is added in the equation to account for the effect of quantum capacitance. The right side of eq.(3) governs how $E_F$ and charge carrier concentrations are correlated with $V_g$ at a given $V_{dirac}$. This correlation along with eqs.(1-2) determines the intrinsic gate transfer characteristics (dashed curves in Fig.2a). The dependence of $V_{dirac}$ on $V_g$ is therefore found from eq.(3) for the graphene device before and after the coating of quantum dots, as shown in Fig.2b. Given the dependence of $E_F$ on $V_g$ in Fig.1d, we can further find the correlation of $V_{dirac}$ with $E_F$, exhibited in Fig.2c (left y axis). Since the concentration ($Q_{ss}$) of localized charges near graphene is given by $Q_{ss} = V_{dirac} C_{ox}/q$ (right y axis in Fig.2c), the density of trap states is further derived from the derivative of $Q_{ss}$ respective to $E_F$ (Fig.2d).

$$q(V_g - V_{dirac}) = E_F + \frac{q^2(n-p)}{C_{ox}} \qquad (3)$$

in which q is the unit charge, $V_g$ is the gate voltage, $V_{dirac}$ is the Dirac point of the intrinsic gate transfer characteristics of graphene, $E_F$ is the Fermi energy level, $C_{ox}$ is the gate oxide capacitance, n and p are the electron and hole concentration in the graphene conduction channel.

Before coating of quantum dots, the concentration of surface charges is written as:

$$Q_{ss} = \int_0^{E_F} D_{it}^{G-O}(E)dE \qquad (4)$$

, in which $E_F$ is the Fermi energy level and $D_{it}^{G-O}(E_F)$ is the density of trap states at graphene and oxide interface.

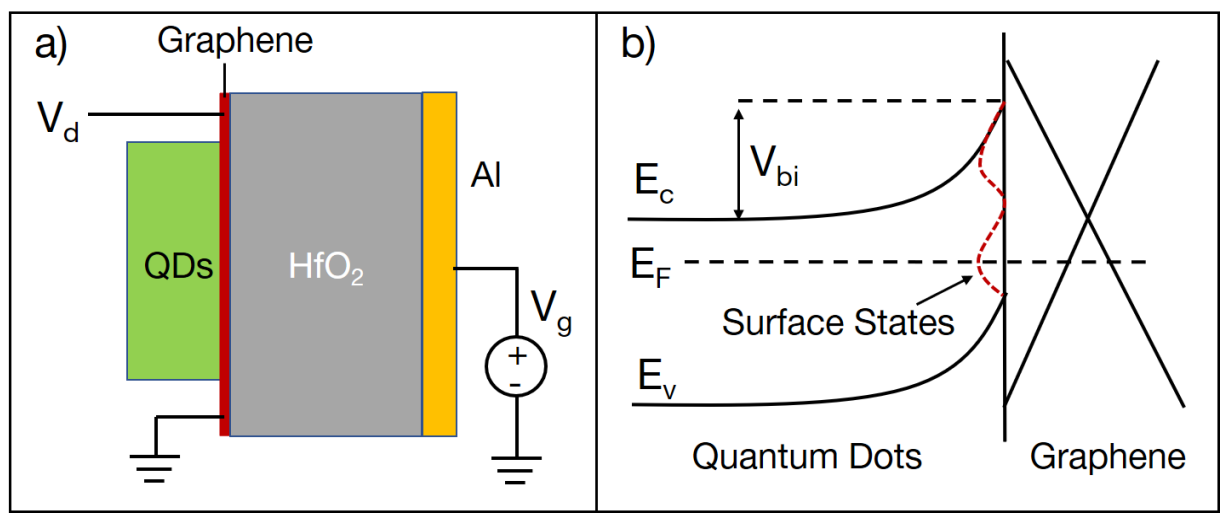

Figure 3 a) Device schematic with quantum dots forming a solid thin film on graphene. b) Energy band diagram of a graphene/QDs device. The energy band in the QDs film bends up towards surface because the work function of the QDs film is smaller than graphene. Electrons will transfer from QDs to graphene after QDs are coated on graphene, which left-shifts the appeared gate transfer characteristics as shown in Fig.1c.

After the graphene device is coated with QDs, the concentration of surface charges is written as eq.(5). The coating of QDs forms a ~30 nm thick compact film. It can be regarded as a continuous solid film self-doped by defects. The device schematic is shown in Fig.3a. Previously we showed that the coating of QDs will left-shift the appeared Dirac point, because the QDs have a smaller work function. Upon coating of QDs, electrons will transfer from QDs into graphene, forming a depletion region in the QDs film. As a result, a built-in potential $V_{bi}$ is established across the depletion region as shown in Fig.3b. A similar scenario was previously observed in literature.[15] In this case, the analytical expression for the effective density of surface states can be written as $D_{it} = D_{it}^{G-O}(E_F) + D_{it}^{G-Q}(E_F) - \frac{W_{dep}N_{eff}}{2V_{bi}}\frac{dV_{bi}}{dE_F}$ by simplifying eq.(5) (see derivation in SI Section 1).

$$Q_{ss} = \int_0^{E_F}[D_{it}^{G-O}(E) + D_{it}^{G-Q}(E)]dE - N_{eff} * W_{dep}(V_{bi}) \tag{5}$$

, where $q$ is the unit charge, $E_F$ is the Fermi energy level, $D_{it}^{G-Q}(E)$ is the density of surface states at the graphene-QDs interface, $N_{eff}$ is the effective self-doping concentration in the QDs film and $W_{dep}$ is the depletion region width that is dependent on the built-in potential $V_{bi}$ in the film.

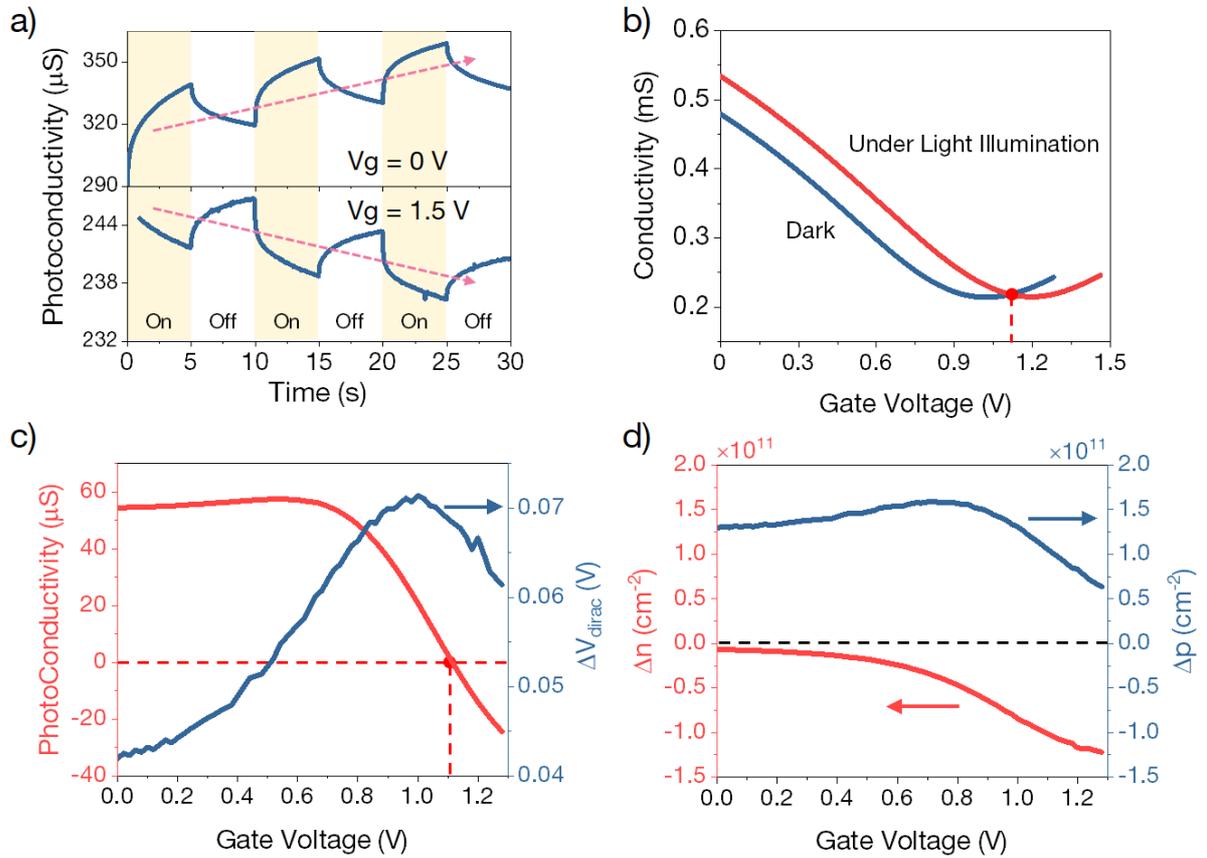

Fig.4 a) Transient photoresponses of graphene/QDs devices upon light illumination that is chopped On/Off. b) Gate transfer characteristics of graphene/QDs device in darkness and under light illumination at the wavelength λ= 532nm and the intensity of 0.003 mW/cm². c) Photoconductivity and intrinsic Dirac point shift at different gate voltage. d) Excess holes Δp and electrons Δn as a function of gate voltage.

Interestingly, the graphene/QDs device responds positively to the light illumination at $V_g$ = 0.6 V and negatively at $V_g$ = 1.5 V as shown in Fig.4a. This is because the light illumination right shifts the gate transfer characteristics (Fig.4b), resulting in positive and negative photoresponses at the left and right side of the appeared Dirac point, respectively. Similar photoresponses were also observed in graphene/QD photodetectors by others.[14] When the gate voltage is far from the appeared Dirac point on the left side, the conductivity of the graphene device is linear with the gate voltage. A right shift in gate transfer characteristics will result in a constant increase in conductivity, i.e. a constant photoconductivity (see red line in Fig.4c). When the gate voltage comes close to the appeared Dirac point (~1V, see Fig.1c), the conductivity deviates from the linear correlation and saturates to the minimum value. The right shift in gate transfer characteristics leads to a reduced photoconductivity. As the gate voltage crosses the red point where the two gate transfer curves intersect, the photoconductivity eventually crosses zero to negative values (Fig.4b and c).

Note that photoconductivity does not come from the light absorption by graphene. Otherwise, the minimal conductivity at appeared Dirac point will increase under light illumination. Previous work showed that the minimal conductivity at the appeared Dirac point decreases under light illumination.[15] It is probably because the light illumination heated up the device.[29] Our device was placed in a temperature-controlled vacuum chamber. The heating effect is minimized. For this reason, photoconductivity must come from the gating effect of photo-induced charge redistribution near graphene. The redistribution of localized charges at a given gate voltage induces a shift in the intrinsic Dirac point $V_{dirac}$. The intrinsic Dirac point shift $\Delta V_{dirac}$ reaches a maximum value (blue curve in Fig.4c) when the gate voltage is around the appeared Dirac point. It is probably because the density of states of graphene is minimized near the Dirac point, as a result of which more of the gate-induced electrons are pumped into surface states, maximizing the intrinsic Dirac point shift $\Delta V_{dirac}$. This intrinsic Dirac point shift $\Delta V_{dirac}$ will change the Fermi energy level and charge carrier concentrations (n and p) following eq.(6) at a fixed gate voltage $V_g$.

$$-q\Delta V_{dirac} = \Delta E_F + \frac{q^2(\Delta n - \Delta p)}{C_{ox}} \qquad (6)$$

Since the photoconductivity comes from the gating effect of photo-induced charge redistribution near graphene, we can analyze the graphene/QDs device under light illumination in the same way as the device was analyzed in darkness ($E_F \sim V_g$ in Fig.1d and $V_{dirac} \sim E_F$ in Fig.2b). It means that $\Delta V_{dirac}$ and $\Delta E_F$ in eq.(6) as a function of $V_g$ can be experimentally found by differentiating the intrinsic Dirac point and Fermi level under light illumination against those in darkness. Following eq.(6), we derived the term $\Delta n - \Delta p$, whereas $\Delta n + \Delta p$ was found from photoconductivity (red curve in Fig.4c) following eq.(1). In the end, the dependence of $\Delta n$ and $\Delta p$ on $V_g$ was calculated separately as shown in Fig.4d. When the gate voltage $V_g$ is at low bias, the graphene is p-type. The photo-induced right shift of the gate transfer characteristics makes the graphene more p-type. As a result, we observed a significant increase in hole concentration and a negligible decrease in electron concentration (electron concentration is already low in p-type graphene). As the gate voltage $V_g$ increases to the right side and graphene becomes n-type, the right-shift of gate transfer characteristics will make graphene less n-type, leading to the dominant decrease in electron concentration and a minor increase in hole concentration, consistent with the experimental observations in Fig.4d.

The photo-induced variation of charge carrier concentrations can be directly measured by photo Hall effect measurements, from which the deduction of $\Delta p$ and $\Delta n$ is dependent on conductance and Hall resistance following eq.(7).[23]

$$\Delta p - \Delta n = \frac{L^2}{e\mu^2 W^2}\left(\sigma^2 \frac{dR_H}{dB} - \sigma_0^2 \frac{dR_{H0}}{dB}\right) \qquad (7)$$

, where $R_{H0}$ and $R_H$ are Hall resistance of the sample in the dark and under light illumination, respectively. The Hall resistances linear with magnetic field in darkness and under light illumination were recorded as shown in SI Section 2. The measured values of $\Delta n+\Delta p$ (from photoconductivity) and $\Delta p-\Delta n$ under different light intensity are exhibited in Fig.5a for $V_g = 0V$ and in Fig.5b for $V_g = 1.5V$. The correlation of $\Delta n$ and $\Delta p$ with the light illumination intensity can be calculated accordingly.

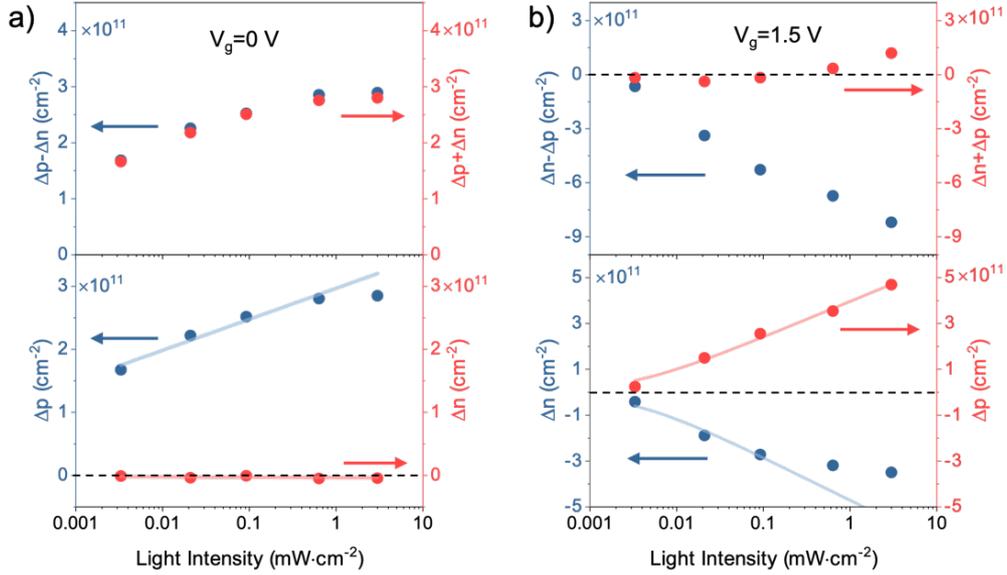

Fig.5 a) Excess electrons and holes found from photo Hall effects and four-probe photoconductance measurements at the gate voltage $V_g = 0V$. b) Excess electrons and holes found from photo Hall effects and four-probe photoconductance measurements at the gate voltage $V_g = 1.5V$.

Interestingly, the correlation of $\Delta n$ and $\Delta p$ with the light illumination intensity can be predicted and fitted theoretically by properly rewriting eq.(6) following the derivation steps below. We first plug eq.(2)and (5) into eq.(6) and then have eq.(8) after reformatting:

$$\Delta V_{bi} = \frac{v(E_F)}{\omega(V_{bi})} \Delta E_F \quad (8)$$

, in which $v(E_F) = \frac{q^2}{C_{ox}}\left[D_{it}^{G-O}(E_F) + D_{it}^{G-Q}(E_F)\right] + 1 + \frac{n_i q^2 \left[J_1'\left(\frac{E_F}{kT}\right) + J_1'\left(-\frac{E_F}{kT}\right)\right]}{kTC_{ox}J_1(0)}$ and $\omega(V_{bi}) = \frac{q^2}{C_{ox}} N_{eff} \cdot \frac{W_{dep}}{2V_{bi}}$

with $J_1'(x) = \frac{dJ_1(x)}{dx}$. All the other parameters have the same physical meanings with previous equations. $v(E_F)$ and $\omega(V_{bi})$ are implicit functions of the gate voltage $V_g$. At a fixed $V_g$, these two parameters are constants unless the light intensity is too strong. Light illumination will shift the built-in potential in the QDs film and Fermi level in graphene. According to the model presented in Fig.3, a photo-induced variation of the built-in potential $V_{bi}$ is the photovoltage $V_{ph}$, that's, $V_{ph} = \Delta V_{bi}$. Our previous work[24] showed that the photovoltage of a surface depletion region can be expressed as in eq.(9).

$$V_{ph} = \frac{\eta kT}{q} \ln\left(\frac{P_{light}}{P_{light}^S} + 1\right) \quad (9)$$

, where $\eta$ is the ideality factor of a floating Schottky junction used to model the surface depletion region, $k$ is the Boltzmann constant, $T$ is the absolute temperature, $q$ is the unit charge, $P_{light}$ is the light illumination intensity and $P_{light}^S$ is the critical light intensity. The critical light intensity is the light intensity at which the photo generation rate in the surface depletion of the QDs film is equal to the thermal generation rate via defects and surface states. The detailed expression of $P_{light}^S$ can be found in ref.[24], which is essentially proportional to the effective recombination rate via defects and surface states. From eq.(8) and (9), we find the correlation of $\Delta E_F$ with light intensity $P_{light}$ as eq.(10).

$$\Delta E_F = kT \frac{\eta\omega(V_{bi})}{qv(E_F)} \ln\left(\frac{P_{light}}{P_{light}^S} + 1\right) \quad (10)$$

, in which $\frac{\eta\omega(V_{bi})}{qv(E_F)}$ is a dimensionless parameter, representing how effectively the light illumination can "photo gate" the graphene. Now we can write the explicit equations for excess electron and hole concentration as eq.(11) by expanding eq.(2) into first order Taylor polynomials of $\Delta E_F$. Eq.(11) shows that excess electron and hole concentration will follow a quasi-logarithmic dependence on light intensity, consistent with the experimental data in Fig.5 except for some deviation in the majority excess carriers at high light intensity ($\Delta p$ at $V_g$ =0 V and $\Delta n$ at $V_g$ =1.5 V).

$$\Delta n = \frac{n_i}{J_1(0)} \frac{\eta\omega(V_{bi})}{qv(E_F)} J_1'\left(\frac{E_F}{kT}\right) \ln\left(\frac{P_{light}}{P_{light}^S} + 1\right); \quad \Delta p = -\frac{n_i}{J_1(0)} \frac{\eta\omega(V_{bi})}{qv(E_F)} J_1'\left(-\frac{E_F}{kT}\right) \ln\left(\frac{P_{light}}{P_{light}^S} + 1\right) \quad (11)$$

We extract the critical light intensity $P_{Light}^S$ and photo gating efficiency $\eta\varpi(V_{bi})/qv(E_F)$ listed in Table I by fitting eq.(11) with the experimental data in Fig.5. When $V_g$ = 0 V, the concentrations of excess minority electrons are too small to reliably fit (third row on right side in Table I). When the gate voltage $V_g$ is switched from 0V to 1.5V, the critical light intensity $P_{Light}^S$ increases by 4 orders of magnitude while the photo gating efficiency is only elevated by one order of magnitude. A 4-order-of-magnitude increase in $P_{Light}^S$ means a dramatic enhancement in the minority carrier recombination rate. To explain this observation, let us recall the scenario when the QDs were applied on graphene. Upon coating of QDs on graphene, electrons were transferred from QDs to graphene due to work function mismatch, which lifted up the Fermi level of graphene (Fig.1c and d). The electron transfer will form a strong inversion depletion region near the QDs film surface, which can electrically passivate the Gr/QDs interface states,[30] resulting in a rather small effective recombination rate and thus a small critical light intensity $P_{Light}^S$. When a positive gate voltage is applied, electrons will be pumped into the graphene from power supply, moving down the graphene energy band (Fig.3b). As a result, the bult-in potential $V_{bi}$ will reduce, pushing the QDs surface depletion region from the strong inversion mode into the depletion mode. More surface states near mid-bandgap will participate the generation-recombination process, which will dramatically increase the minority carrier recombination rate[20] and also $P_{Light}^S$, consistent with the fitting results in Table I.

The increase of the photo gating efficiency $\eta\varpi(V_{bi})/qv(E_F)$ can be understood by examining how $\varpi(V_{bi})$ and $v(E_F)$ change when the gate increases from 0V to 1.5V. Fig.2d exhibits the effective density of trap states near graphene which is significantly reduced by the coating of QDs. On the left end ($V_g \approx 0$ V) and right end ($V_g \approx 1.5$ V), the reduction in the effective density of trap states is $0.24 \times 10^{14}$ cm$^{-2}$/eV and $1.8 \times 10^{14}$ cm$^{-2}$/eV, respectively. It can be seen from eq.(4) and (5) that this reduction is mainly contributed by $\varpi(V_{bi})C_{ox}/q$, meaning that $\varpi(V_{bi})$ is increased by a factor of 8 ($\approx 1.8/0.24$). For $v(E_F)$, there are three terms. The first term, i.e., the trap state density at the Gr/SiO$_2$ interface are roughly the same at Vg $\approx$ 0V and 1.5V (see Fig.2d). It is reasonable to assume that the trap state density at the Gr/QDs interface (the second term) is also comparable at these two gate voltages. The last term in $v(E_F)$ is associated with the first derivative of two J functions. A simple calculation show that this term becomes smaller by a factor of ~ 1.5 after the gate voltage is switched from 0 V to 1.5 V. The summation of these three terms indicates that $v(E_F)$ does not change very much within this gate sweeping range. In short, the photo gating efficiency $\eta\varpi(V_{bi})/qv(E_F)$ should at least increase by a factor of 8 when $V_g$ increases from 0 V to 1.5 V. This is largely consistent with our fitting results in Table I.

Table I. Parameters extracted by fitting eq.(11) with experimental data.

| Gate voltage | $E_F$ (eV) | Correlations | $P^s_{Light}$ (μW/cm$^2$) | $\eta\varpi(V_{bi})/qv(E_F)$ |
|---|---|---|---|---|
| $V_g = 0$ V | -0.0576 | $\Delta p \sim P_{Light}$ | $(0.93 \pm 1.2) \times 10^{-3}$ | $(-3.1 \pm 0.4) \times 10^{-2}$ |
| | | $\Delta n \sim P_{Light}$ | - | - |
| $V_g = 1.5$ V | 0.0185 | $\Delta p \sim P_{Light}$ | $2.9 \pm 1.2$ | $-0.57 \pm 0.05$ |
| | | $\Delta n \sim P_{Light}$ | $3.0 \pm 2.8$ | $-0.24 \pm 0.08$ |

Given Δn and Δp, it is not difficult to find the photoconductance that is a function of light intensity as shown in eq.(12).

$$\Delta\sigma = q \cdot \mu \cdot \frac{W}{L}(\Delta n + \Delta p) = q\frac{\mu W n_i}{L}\frac{E_F}{J_1(0)kT}\frac{\eta\omega(V_{bi})}{qv(E_F)}\ln\left(\frac{P_{light}}{P^s_{light}} + 1\right) \tag{12}$$

The photo gain is expressed as eq.(13).

$$G = \frac{I_{ph}/q}{P_{light}A_{proj}/\hbar\omega} = G_{max}\frac{P^s_{light}}{P_{light}}\ln\left(\frac{P_{light}}{P^s_{light}} + 1\right) \tag{13}$$

, in which $G_{max} = \frac{\hbar\omega\mu V_{ds} n_i}{L^2 P^s_{light}}\frac{E_F}{J_1(0)kT}\frac{\eta\omega(V_{bi})}{qv(E_F)}$. Since $P^s_{Light}$ is inversely proportional to the effective minority recombination lifetime τ (in QDs instead of graphene), the maximum photo gain G$_{max}$ will be proportional to the ratio of τ to the transit time $\tau_t = \frac{L^2}{\mu V_{ds}}$, similar to the classical gain theory in this aspect. However, unlike the classical photo gain theory in which the gain follows a simple equation of $G = \frac{\tau}{\tau_t}(1 + \frac{\mu_n}{\mu_p})$, our gain equation predicts that the photo gain is also dependent on light intensity, density of trap states and energy band structure of underlying materials (J functions for graphene), similar to what we previously found for nanowire photoconductors.[24]

As expected, eqs.(12) and (13) fit well with the experimental data at V$_g$ = 0V in Fig.6a except for the data at high light intensity. The extracted parameters are summarized in Table II and comparable to those in Table I for the case of V$_g$ = 0V. The maximum photo gain G$_{max}$ is 1.1 x 10$^7$. Surprisingly, these equations do not fit well with the photoresponses at V$_g$ =1.5V in Fig.6b. As the light intensity increases, the negative photoconductivity first decreases to a minimal value when the gate transfer characteristics under illumination (red curve in the inset of Fig.6b top panel) right-shifts by photo gating effect till its Dirac point reaches 1.5V (the applied gate voltage V$_g$). The photoconductivity further increases across zero to positive values when the gate transfer characteristics under illumination and in darkness intersect at V$_g$ = 1.5V.

To understand this nonlinear phenomenon, let us examine carefully Δn and Δp in Fig.5b. We see that Δn and Δp are opposite in sign and comparable in magnitude with some nonlinearity. The photoconductance is proportional to the difference of their magnitude (Δp + Δn = |Δp| - |Δn|) which cancels out the main linear components, leaving the minor nonlinear components to dominate the photoconductance (resulting in much smaller photoconductivities in Fig.6b compared to Fig.6a). As a result, high order Taylor polynomials should be added into eq.(11) so that the nonlinear dependence on light intensity can be caught for the photoconductance and photo gain. The explicit photoconductivity and photo gain equations with 2nd order Taylor polynomial are presented in the SI Section 3. These equations can fit the experimental well in Fig.6b. Parameters extracted from the fittings are presented in Table II. The extracted critical light intensity $P^s_{Light}$ and photo gating efficiency for the case of V$_g$ =1.5V are close to the values found in Table I, validating our model and analytical equations. When the illumination intensity is reduced, the 2$^{nd}$ order Taylor polynomial rapidly becomes negligible in comparison with the first order terms. As a result, the negative gain increases (more negative) and eventually saturates to a maximum gain of - 4.2 x 10⁴ governed by eq.(13).

Table II. Parameters extracted by fitting theoretical equations with experimental data in Fig.6.

| Gate Voltage | $E_F$ (eV) | $P_{Light}^{S}$ (μW/cm²) | $\eta\varpi(V_{bi})/qv(E_F)$ | $G_{max}$ |
|---|---|---|---|---|
| $V_g$ = 0 V | -0.0576 | $(1.2 \pm 1.3) \times 10^{-3}$ | $(-4.6 \pm 0.5) \times 10^{-2}$ | $(1.1 \pm 0.07) \times 10^7$ |
| $V_g$ =1.5 V | 0.01846 | $1.6 \pm 1.3$ | $-0.27 \pm 0.03$ | $-(4.2 \pm 1.7) \times 10^4$ |

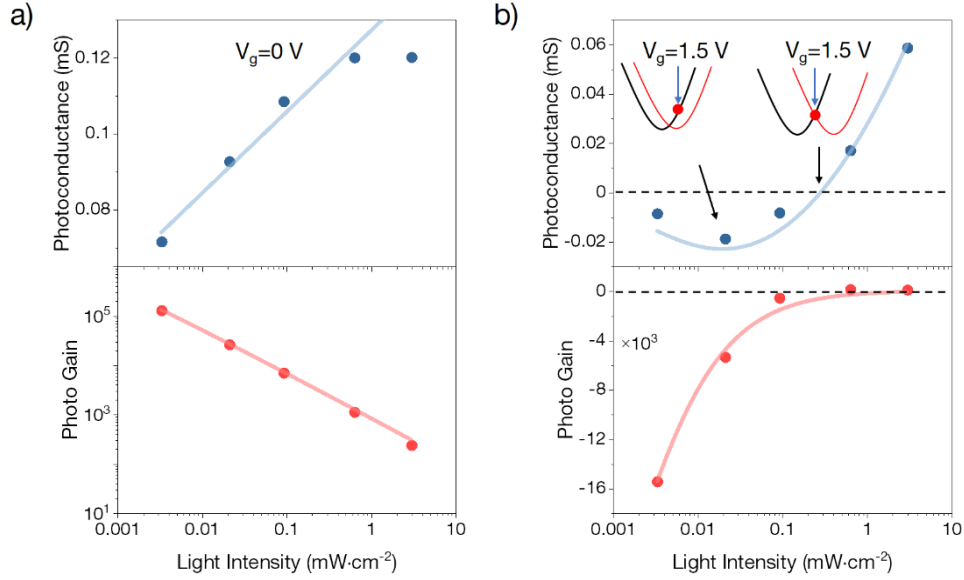

Fig.6 a) Photoconductivity and photo gain at $V_g$ = 0V. Dots are experimental data and solid lines in top and bottom panel are fitting lines of theoretical equations eq.(12) and (13), respectively. b) Photoconductivity and photo gain at $V_g$ = 0V. Dots are experimental data and solid lines in top and bottom panel are fitting lines of theoretical equations eq.(S6) and eq.(S7) in SI Section 3, respectively.

**Conclusions**

In this work, we managed to derive explicit photogain equations for hybrid QD-graphene photodetectors. The equations fit well the positive and negative photoresponses of the graphene device. The physical parameters extracted from the fitting are largely consistent with our quantitative analysis. These gain equations can be used to design and predict the photoresponses of similar hybrid graphene-quantum-dot photodetectors if the properties of QDs and the device fabrication is well controlled. More importantly, the way we derived these explicit gain equations may be readily applied to derive explicit gain equations for all 2D semiconducting photoconductors.

**Experimental Section**

**Device Fabrication** Graphene/QDs photodetectors were fabricated on a highly doped Si wafer with a 2 μm thick SiO₂ on the top. A 150 nm thick aluminum gate electrode was first deposited on the Si/SiO₂ wafer by photolithography (Mask Aligner MA6) and thermal evaporation (Angstrom, Canada). After liftoff process and cleaning, a layer of HfO₂ 30 nm thick was then grown on the sample surface (covering the Al gate electrode) at 250 °C by plasma enhanced atomic layer deposition (PEALD, Beneq). Next, we performed a second time of photolithography, thermal evaporation and liftoff to $(NH_4)_2S_2O_8$ solution pattern five Au/Cr electrodes (100nm/15nm) that are properly aligned to the Al gate electrode. Later on, a monolayer of graphene sheet was transferred from Cu foil to the sample surface in contact with the five electrodes. During graphene transfer process, polymethylemethacrylate (PMMA) was first spin coated onto the copper foil with graphene synthesized by chemical vapor deposition (purchased from ACS Materials). The sample was then immersed in 0.5 mol/L $(NH_4)_2S_2O_8$ solution to remove the copper foil. After the copper foil was dissolved, the PMMA/Graphene membrane was left floating on the solution. The $(NH_4)_2S_2O_8$ solution was slowly diluted by de-ionized water. The PMMA/Graphene membrane was eventually floating on de-ionized wafer and the wafer with pre-fabricated electrodes was subsequently immersed in it to "catch" the floating PMMA/Graphene membrane. The as-obtained samples were further dried in ambient for at least one day. The PMMA was removed by immersing the sample in acetone for 30 min, followed by washing with acetone for at least 3 times to remove residual PMMA. The sample was then cleaned in isopropanol and de-ionized water. To pattern the graphene sheet into a Hall bar geometry, we employed photo lithography to pattern the spin-coated positive photoresist (S1813, Microchem) that protected part of the graphene sheet. After oxidation plasma (PE-100 Plasma Etch Benchtop System) was used to remove the graphene unprotected by photoresist, the photoresist was removed by immersing the sample in acetone for 24 hours.

**QDs Coating** PbS QDs were first immersed in 2% ethanedithiol (EDT) in acetonitrile (ACN) solution to grow EDT ligands on PbS QDs and increase the carrier mobility in PbS QDs. Secondly, PbS QDs with ligands were dispersed in toluene (30 mg/mL) and spin coated on the graphene at the speed of 2000 rpm/min. After dried at room temperature for more than 10 seconds, 2% EDT in ACN solution was spin-coated on the sample at the same speed. Finally, mixture solution of ACN and toluene (1:1, v/v) was used for cleaning by the same spin-coating process twice. In the end, a layer of QDs-ligands approximately 50 nm thick was formed on the graphene.

**Electrical Measurements** The samples were placed in a physical property measurement system (PMMS, Evercool-II) in vacuum at a controlled temperature of 300K. One high accuracy picoammeter (Keithley 2636B) and two digital sourcemeters (Keithley 2400) were connected to PPMS system to supply voltage and measure voltage and current. The system is controlled via GPIB by Labview scripts. Keithley 2636B was applied to measure the Hall resistance. One Keithley 2400 was applied to measure the electronic properties and the other one to supply the gate voltage. Hall measurements were performed on the graphene/QDs photoconductors in dark environment to extract the information on charge carriers. To measure photo Hall effect resistance, a green LED with peak light intensity at the wavelength of 530nm was placed in the chamber of PPMS systems. The light intensity was modulated by adjusting the input power in the LED. The light intensity was calibrated by a commercial photodetector (G10863, Hamamatsu).

**Acknowledgements**
The work is financially supported by the special-key project of the "Innovative Research Plan", Shanghai Municipality Bureau of Education (2019-01-07-00-02-E00075), the Key R&D Program of Zhejiang Province (2019C01155) and National Science Foundation of China (NSFC) (61874072). The devices were fabricated at the center for Advanced Electronic Materials and Devices (AEMD), and photo Hall measurements were conducted at the Instrumental Analysis Center (IAC), Shanghai Jiao Tong University. We thank Prof. Zhenhua Ni for providing us some of the graphene samples and Mr. Shubin Su for the assistance of plasma oxidation, passivation and graphene transfer.

**References**

1. Geim AK, Novoselov KS. The rise of graphene. *Nature Materials* **6**, 183-191 (2007).
2. Bolotin KI, *et al.* Ultrahigh electron mobility in suspended graphene. *Solid State Commun* **146**, 351-355 (2008).
3. Morozov SV, *et al.* Giant Intrinsic Carrier Mobilities in Graphene and Its Bilayer. *Phys Rev Lett* **100**, 016602 (2008).
4. Chen J-H, Jang C, Xiao S, Ishigami M, Fuhrer MS. Intrinsic and extrinsic performance limits of graphene devices on SiO2. *Nature Nanotechnology* **3**, 206-209 (2008).
5. Novoselov KS, *et al.* Two-dimensional gas of massless Dirac fermions in graphene. *Nature* **438**, 197-200 (2005).
6. Liu C-H, Chang Y-C, Norris TB, Zhong Z. Graphene photodetectors with ultra-broadband and high responsivity at room temperature. *Nature Nanotechnology* **9**, 273-278 (2014).
7. Yuan S, *et al.* Room Temperature Graphene Mid-Infrared Bolometer with a Broad Operational Wavelength Range. *Acs Photonics* **7**, 1206-1215 (2020).
8. Zhang BY, *et al.* Broadband high photoresponse from pure monolayer graphene photodetector. *Nature Communications* **4**, 1811 (2013).
9. Yu X, *et al.* A high performance, visible to mid-infrared photodetector based on graphene nanoribbons passivated with HfO2. *Nanoscale* **8**, 327-332 (2016).
10. Nair RR, *et al.* Fine Structure Constant Defines Visual Transparency of Graphene. *Science* **320**, 1308-1308 (2008).
11. Xia F, *et al.* Photocurrent Imaging and Efficient Photon Detection in a Graphene Transistor. *Nano Letters* **9**, 1039-1044 (2009).
12. Xia F, Mueller T, Lin Y-m, Valdes-Garcia A, Avouris P. Ultrafast graphene photodetector. *Nature Nanotechnology* **4**, 839-843 (2009).
13. Gan X, *et al.* Chip-integrated ultrafast graphene photodetector with high responsivity. *Nature Photonics* **7**, 883-887 (2013).
14. Zheng L, *et al.* Ambipolar Graphene–Quantum Dot Phototransistors with CMOS Compatibility. *Advanced Optical Materials* **6**, 1800985 (2018).
15. Konstantatos G, *et al.* Hybrid graphene–quantum dot phototransistors with ultrahigh gain. *Nature Nanotechnology* **7**, 363-368 (2012).
16. Turyanska L, *et al.* Ligand-Induced Control of Photoconductive Gain and Doping in a Hybrid Graphene–Quantum Dot Transistor. *Advanced Electronic Materials* **1**, 1500062 (2015).
17. Konstantatos G, *et al.* Ultrasensitive solution-cast quantum dot photodetectors. *Nature* **442**, 180-183 (2006).
18. Konstantatos G, Sargent EH. Colloidal quantum dot photodetectors. *Infrared Physics & Technology* **54**, 278-282 (2011).
19. Konstantatos G, Sargent EH. Nanostructured materials for photon detection. *Nature Nanotechnology* **5**, 391-400 (2010).
20. Neamen DA. *Semiconductor physics and devices: basic principles*. New York, NY: McGraw-Hill (2012).
21. Dan Y, Zhao X, Chen K, Mesli A. A Photoconductor Intrinsically Has No Gain. *Acs Photonics* **5**, 4111-4116 (2018).
22. Petritz RL. Theory of Photoconductivity in Semiconductor Films. *Physical Review* **104**, 1508-1516 (1956).
23. Chen K, Zhao X, Mesli A, He Y, Dan Y. Dynamics of Charge Carriers in Silicon Nanowire

Photoconductors Revealed by Photo Hall Effect Measurements. *ACS Nano* **12**, 3436-3441 (2018).

24. He J, Chen K, Huang C, Wang X, He Y, Dan Y. Explicit Gain Equations for Single Crystalline Photoconductors. *ACS Nano* **14**, 3405-3413 (2020).

25. Dan Y, Seo K, Takei K, Meza JH, Javey A, Crozier KB. Dramatic Reduction of Surface Recombination by in Situ Surface Passivation of Silicon Nanowires. *Nano Letters* **11**, 2527-2532 (2011).

26. Grotevent MJ, *et al.* Nanoprinted Quantum Dot–Graphene Photodetectors. *Advanced Optical Materials* **7**, 1900019 (2019).

27. Fang T, Konar A, Xing H, Jena D. Carrier statistics and quantum capacitance of graphene sheets and ribbons. *Applied Physics Letters* **91**, 092109 (2007).

28. Zebrev G, Melnik E, Tselykovskiy A. Interface traps in graphene field effect devices: extraction methods and influence on characteristics. *arXiv preprint arXiv:14055766*,  (2014).

29. Ma N, Jena D. Carrier statistics and quantum capacitance effects on mobility extraction in two-dimensional crystal semiconductor field-effect transistors. *2d Mater* **2**, 015003 (2015).

30. He J, *et al.* High-Efficiency Silicon/Organic Heterojunction Solar Cells with Improved Junction Quality and Interface Passivation. *ACS Nano* **10**, 11525-11531 (2016).